\documentclass[runningheads,a4paper]{llncs}
\usepackage{natbib}
\setcitestyle{round}
\usepackage{graphicx}
\usepackage{color}
\usepackage{amsmath}
\usepackage{url}

\begin{document}
\sloppy

\mainmatter

\title{Molecular dynamics simulations of chemically modified ribonucleotides}
\titlerunning{Molecular dynamics simulations of chemically modified RNA}

\author{Valerio Piomponi\inst{1} \and Mattia Bernetti\inst{2} \and Giovanni Bussi\inst{1}}
\authorrunning{Molecular dynamics simulations of chemically modified RNA}

\institute{Scuola Internazionale Superiore di Studi Avanzati - SISSA,\\ via Bonomea 265, 34136 Trieste, Italy
\and
Computational and Chemical Biology, Italian Institute of Technology,\\ 16152 Genova, Italy and
Department of Pharmacy and Biotechnology,\\ Alma Mater Studiorum - University of Bologna, 40126 Bologna, Italy
}

\maketitle
\begin{abstract}
Post-transcriptional modifications are crucial for RNA function, with roles ranging from the stabilization of functional RNA structures to modulation of RNA--protein interactions. Additionally, artificially modified RNAs have been suggested as optimal oligonucleotides for therapeutic purposes. The impact of chemical modifications on secondary structure has been rationalized for some of the most common modifications. However, the characterization of how the modifications affect the three-dimensional RNA structure and dynamics and its capability to bind proteins is still highly challenging.
Molecular dynamics simulations, coupled with enhanced sampling methods and integration of experimental data, provide a direct access to RNA structural dynamics. In the context of RNA chemical modifications, alchemical simulations where a wild type nucleotide is converted to a modified one are particularly common. In this Chapter, we review recent molecular dynamics studies of modified ribonucleotides. We discuss the technical aspects of the reviewed works, including the employed force fields, enhanced sampling methods, and alchemical methods, in a way that is accessible to experimentalists. Finally, we provide our perspective on this quickly growing field of research. The goal of this Chapter is to provide a guide for experimentalists to understand molecular dynamics works and, at the same time, give to molecular dynamics experts a solid review of published articles that will be a useful starting point for new research.
\end{abstract}

\newpage

List of abbreviations: \vspace{0.2cm}

\begin{tabular}{ll}
A           & adenosine \\
C           & cytidine \\
G           & guanosine \\
i$^6$A      & N6-isopentenyladenosine  \\
I           & inosine \\
LNA         & locked nucleic acid \\
m$^1$A      & N1-methyladenosine \\
m$^1$G      & N1-methylguanosine \\
m$^2$G      & N2-methylguanosine \\
m$^2$$_2$G  & N2-dimethylguanosine \\
m$^6$A      & N6-methyladenosine \\
m$^6$$_6$A  & N6-dimethyladenosine \\
MD          & molecular dynamics \\
mRNA        & messenger RNA \\
NMR         & nuclear magnetic resonance \\
$\Psi$      & pseudouridine \\
PT          & Phosphorothioate \\
rRNA        & ribosomal RNA \\
s$^2$U      & 2-thiouridine \\
s$^4$U      & s42-thiouridine \\
siRNA       & small interfering RNAs \\
tRNA        & transfer RNA \\
U           & uridine          \\
\end{tabular}
\newpage

\section{Introduction}

\begin{figure}
\begin{center}
\includegraphics[width=0.8\textwidth]{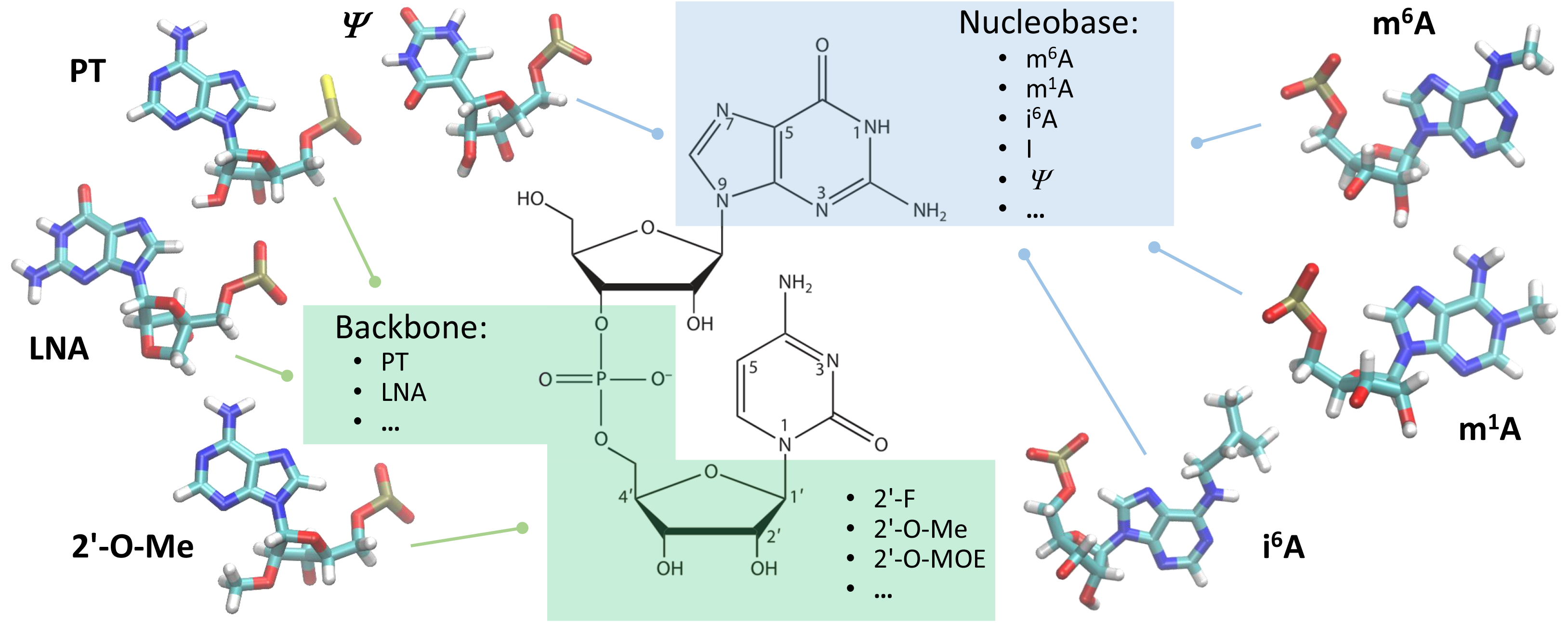}
\end{center}
\caption{
\small
Schematic representing some of the nucleotide modifications that are discussed in this Chapter.
\label{fig-modifications}
}
\end{figure}

RNA molecules are sequences of four common nucleotides:
adenosine (A), uridine (U), cytidine (C), and guanosine (G).
However, a number of naturally occurring or artificially synthesized nucleotides can be incorporated as well (see Fig.~\ref{fig-modifications}).
Naturally occurring modifications are often
chemical marks on cellular RNA, are regulated and recognized by proteins known as \emph{writers} and \emph{readers}, respectively,
and are usually referred to as \emph{post-transcriptional} modifications.
After the first modification  was discovered \citep{davis1957}, more than a hundred of them have been identified.
Transfer RNAs (tRNAs) are known to be heavily modified \citep{nachtergaele2017emerging}, with a variety of modifications found both in the anticodon region and in the tRNA-body region \citep{ramos2019,suzuki2021}.
Ribosomal RNA (rRNA) is also extensively edited after transcription \citep{jiang2016post}.
Recent technical advances revealed widespread modifications also on messenger RNAs (mRNAs) \citep{gilbert2016}.
In general, post-transcriptional modifications have two roles: (i) they allow correct folding of noncoding RNAs (e.g., tRNA and rRNA) into their functional structure and
(ii) they affect the target specificity of RNA--RNA and RNA--protein interactions.
In addition to naturally occurring modifications, a number of artificially modified nucleotides have been studied, mostly in the context of oligonucleotide design \citep{wan2016medicinal}. Synthetic oligonucleotides hold great potential as innovative therapeutic strategies, including small interfering RNAs (siRNAs), antisense oligonucleotides, microRNAs, and aptamers. However, intrinsic limitations in terms of instability, immunogenicity, and poor pharmacokinetic properties hamper their use. Therefore, artificial modifications of RNA nucleotides were explored to overcome these limitations and optimize the oligonucleotide properties.
A remarkable example in this respect, though through artificial insertion of a natural modification, is pseudo-uridine ($\Psi$) in mRNA-vaccine technology \citep{kariko2008incorporation}.

Although the research on RNA modifications has been exponentially increasing in the past years, computational studies on modified RNAs are still limited, even in the relatively simpler context of secondary structure prediction \citep{tanzer2019}.
Additional complexity is present in the study of the impact of modifications on tertiary structure.
Models for tertiary structure predictions are typically trained on available structural datasets \citep{parisien2008mc,townshend2021geometric},
and the statistics available on modified nucleotides is scarce. Furthermore, methods trained on static structures give limited access to structural dynamics. In this respect,
molecular dynamics (MD) simulations \citep{dror2012biomolecular,sponer2018rna} are a promising tool since (a) they give direct access to dynamics and (b) they are grounded in physics-based models, which could possibly be capable to describe systems for which the amount of reference experimental structures is limited.

The aim of this Chapter is to review recent applications of MD simulations to the study of chemically modified ribonucleotides.
In particular, we will first describe the basic principles of MD simulations,
including advanced methods to enhance sampling and compute mutation free energies.
Then, we will review selected applications of MD to the study of the effect of modifications on RNA
structural dynamics and RNA recognition. Finally, we will provide our perspective on the field.

\section{Molecular dynamics simulations}

\subsection{Standard molecular dynamics simulations}

Molecular dynamics simulations are a natural tool to characterize RNA structural dynamics \citep{sponer2018rna}.
A key ingredient of any molecular dynamics (MD) simulations is the employed force field, that is a function that computes the
forces on the atoms given their positions.
Since evaluating the force field is the computational bottleneck of
MD simulation,
its functional form has to be chosen with compromises, so as to be accurate enough to describe the relevant chemistry
but not too expensive. The functional form of the commonly used AMBER \citep{cornell1995second} force field is the following one:

\begin{multline}
E= \sum_{bonds} \frac{1}{2}k_b(r-r_0)^2 + \sum_{angles} \frac{1}{2}k_a(a-a_0)^2 + \sum_{torsions} \sum_n\frac{V_n}{2}(1+\cos(n\phi-\delta)) +
\\ \sum_{LJ}4 \epsilon_{ij}\left(\left(\frac{\sigma_{ij}}{r_{ij}}\right)^{12} - \left(\frac{\sigma_{ij}}{r_{ij}}\right)^6\right) + \sum_{electrostatics} \frac{q_iq_j}{r_{ij}}
\label{eq:ff}
\end{multline}
Here, $k_b$, $k_a$, and $V_n$ control the so-called bonded interactions (bonds, angles, and torsional angles, respectively),
$\sigma$ and $\epsilon$ control Lennard-Jones potentials, and charges $q$ control electrostatic interactions.

The parameters of the force field are heavily system dependent and are derived using a mixture of accurate quantum chemical calculations and of experimental data
(see \cite{frohlking2020toward} for a recent review).
The two main families of force fields used for nucleic acids are AMBER \citep{cornell1995second} and CHARMM \citep{mackerell1995all},
both of which have evolved in multiple revised versions during the past decades.

\subsection{Force fields for chemically modified nucleotides}

Specific force-field parameters should be derived for each type of modification.
The AMBER family of force fields offers a well-defined recipe for arbitrary molecules.
In particular, charges are obtained by fitting the electrostatic potential, and torsional parameters by fitting the energy profiles associated to bond rotations.
For the CHARMM force field, the procedure is more complex and targets both quantum mechanical data on nucleosides and experimental data on nucleosides or oligonucleosides.
Luckily, parameters obtained using these procedures have been published
for approximately 100 naturally occurring modified nucleotides,
both in the AMBER \citep{aduri2007amber} and in the CHARMM \citep{xu2016additive} frameworks.

In \cite{aduri2007amber},  parameters were
validated performing standard MD
simulations of a tRNA containing a fraction of the modifications for which parameters were reported.
These force-field parameters have been used in several later MD simulations using the AMBER force field (see below),
and are also used in modelling tools (see, e.g., \cite{stasiewicz2019qrnas}).

\cite{xu2016additive} provided force-field parameters for 112 modified nucleotides and tested in detail 13 of them.
Tautomers and protonation variants have also been included.
The charge-fitting strategy aims at reproducing interactions with water and correct dipole moments.
 Torsions were fitted computing potential energy surfaces with quantum mechanical methods.
Simulations of nucleosides and trinucleotides were compared with nuclear-magnetic-resonance (NMR) data, when available.
These force-field parameters have been used in several later MD simulations using the CHARMM force field (see below).

In addition to these two works,
it is relevant to mention that for many of the applications discussed below the authors
developed and tested new sets of force-field parameters specific for a single or a few modifications.

\subsection{Enhanced sampling methods}

RNA molecules are often characterized by conformational ensembles composed of multiple partly heterogeneous structures or substates
that are relevant for function \citep{ganser2019roles}.
MD simulations can access at most the multi-microsecond timescale with current resources.
Changes in tertiary interactions and base-pairing pattern cannot thus be
directly simulated with MD. To circumvent this problem, enhanced sampling methods can be used.

Enhanced sampling methods \citep{mlynsky2018exploring,henin2022enhanced} are roughly classified in two categories. One category includes methods based on heating
the system so as to accelerate the exploration of the conformational space.
Representatives of these methods are parallel tempering,
also known as temperature replica exchange \citep{sugita1999replica}, and solute tempering \citep{wang2011replica}.
These methods are typically very expensive and can thus be fruitfully applied only for sampling small oligomers.
The other category includes methods based on adding biasing forces on specifically chosen degrees of freedom, or collective variables.
Representatives of these methods are umbrella sampling
\citep{torrie1977nonphysical}, often performed combining multiple windows \citep{kumar1992weighted} so as to progressively convert the system from
an initial conformation to a final one, and metadynamics \citep{laio2002escaping}. These methods can be used to accelerate
relevant events if sufficient prior information about the  process is given.

\subsection{Alchemical methods}

\begin{figure}
\begin{center}
\includegraphics[width=0.8\textwidth]{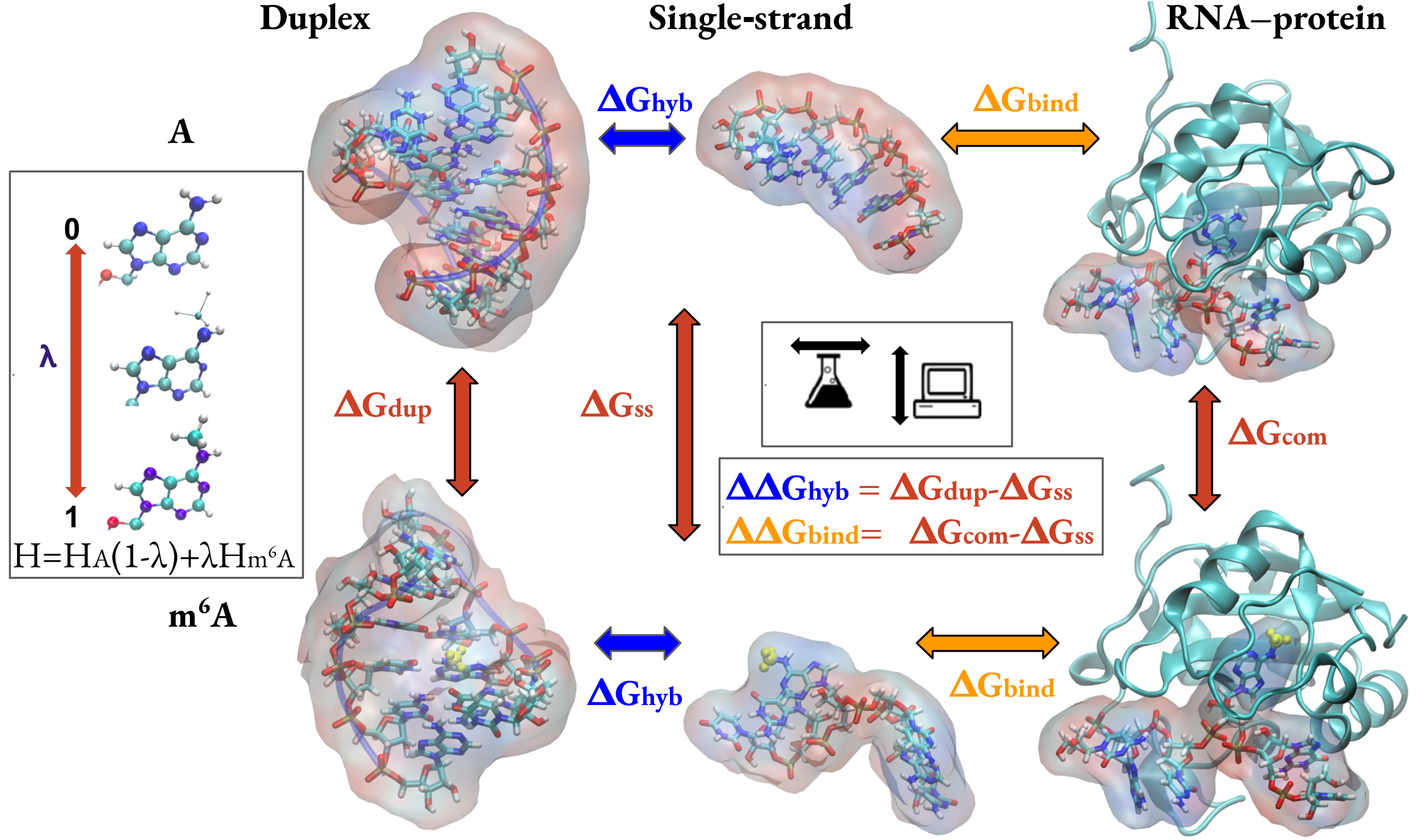}
\end{center}
\caption{
\small
Schematic representation of alchemical simulation protocols. Transformation
A$\leftrightarrow$m$^6$A (vertical red arrows) can be used to compute the free-energy change
in different structural contexts (duplex, $\Delta G_{\textrm{dup}}$, single strand, $\Delta G_{\textrm{ss}}$, and RNA--protein complex, $\Delta G_{\textrm{com}}$).
The Hamiltonian function ($H$) is interpolated between the two physical systems.
Experiments (horizontal blue and orange arrows) report hybridization free energies ($\Delta G_{\textrm{hyb}}$)
and free energies of binding ($\Delta G_{\textrm{bind}}$).
$\Delta \Delta G$s can be directly compared between simulation and experiment. E.g.,
$\Delta G_{\textrm{dup}} - \Delta G_{\textrm{ss}} = \Delta G_{\textrm{hyb,m}^6\textrm{A}} - \Delta G_{\textrm{hyb,A}}$
\label{fig-alchemistry}
}
\end{figure}

Alchemical methods allow to simulate trajectories where molecular species are mutated to different ones, and the free-energy
associated to the transformation can be computed \citep{mey2020best}. %
Most MD code support these methods, but setting up the simulations
is usually more complex than for standard MD. The intermediate states might be simulated independently  of each other
or with a more robust replica-exchange procedure \citep{meng2011computing}.
Simulations should be then repeated in different structural contexts.
For instance,
the conversion between (unmodified) A and (modified) m$^6$A can be performed in a single strand
and in a duplex. The difference between the free-energy changes computed in the two contexts
corresponds to the stabilization of the duplex resulting from the additional methylation
(see Fig.~\ref{fig-alchemistry}).
Similarly, the impact of the modification on the affinity between the studied RNA and
a protein can be estimated.

Results of alchemical simulations should be judged with care.
Specifically, if there are slow degrees of freedom coupled with the alchemical change,
the result might be affected by artifacts.
A typically difficult situation
arises when one or both the alchemical states correspond to flexible conformations whose extensive sampling is difficult.
A possible way to alleviate this problem consists in combining alchemical simulations with enhanced sampling methods,
as done for instance in alchemical metadynamics \citep{hsu2022adding}.

\section{Applications}

\subsection{Validation and fitting of force fields against experimental data}
\label{sec-validation}

Quantitative validations are crucial to assess the capability of force-field parameters to generate structural ensembles compatible with experiments.

\cite{deb2014conformational} showed that the Aduri parametrization is not able to reproduce
substate populations for a number of modified uridines. Later, they provided a reparametrization for
$\Psi$, s$^2$U  and s$^4$U \citep{deb2016reparameterizations}, and discussed how changes in the Lennard-Jones parameters
could affect the relative population of different sugar conformations.
\cite{dutta2020molecular} later confirmed these parameters to be transferable to other modified uridines.
More recently, \cite{dutta2022data} reparametrized charges for $\Psi$ and three different methylated versions of $\Psi$,
obtaining parameters that were then validated simulating single-stranded oligonucletides, obtaining conformational and hydration properties in agreement with
experimental data.

\cite{hurst2021deciphering} have validated Aduri parameters for  m$^6$A using alchemical simulations, confirming their capability to
reproduce melting experiments. However, a later work from our group \citep{piomponi2022molecular}
using a similar protocol on a larger validation set showed that modifications to the charges are necessary to simultaneously
reproduce thermodynamic data and \emph{syn/anti} balance for the methyl group.

\subsection{Effects of post-transcriptional modifications on RNA structure and dynamics}

Here we review MD simulation studies aimed at elucidating the impact of post-transcriptional modifications on the structure and dynamics of
tRNAs, rRNAs, and other systems.

\subsubsection{Molecular simulations of anti-codon stem loops and of entire tRNAs.}

\begin{figure}
\begin{center}
\includegraphics[width=0.8\textwidth]{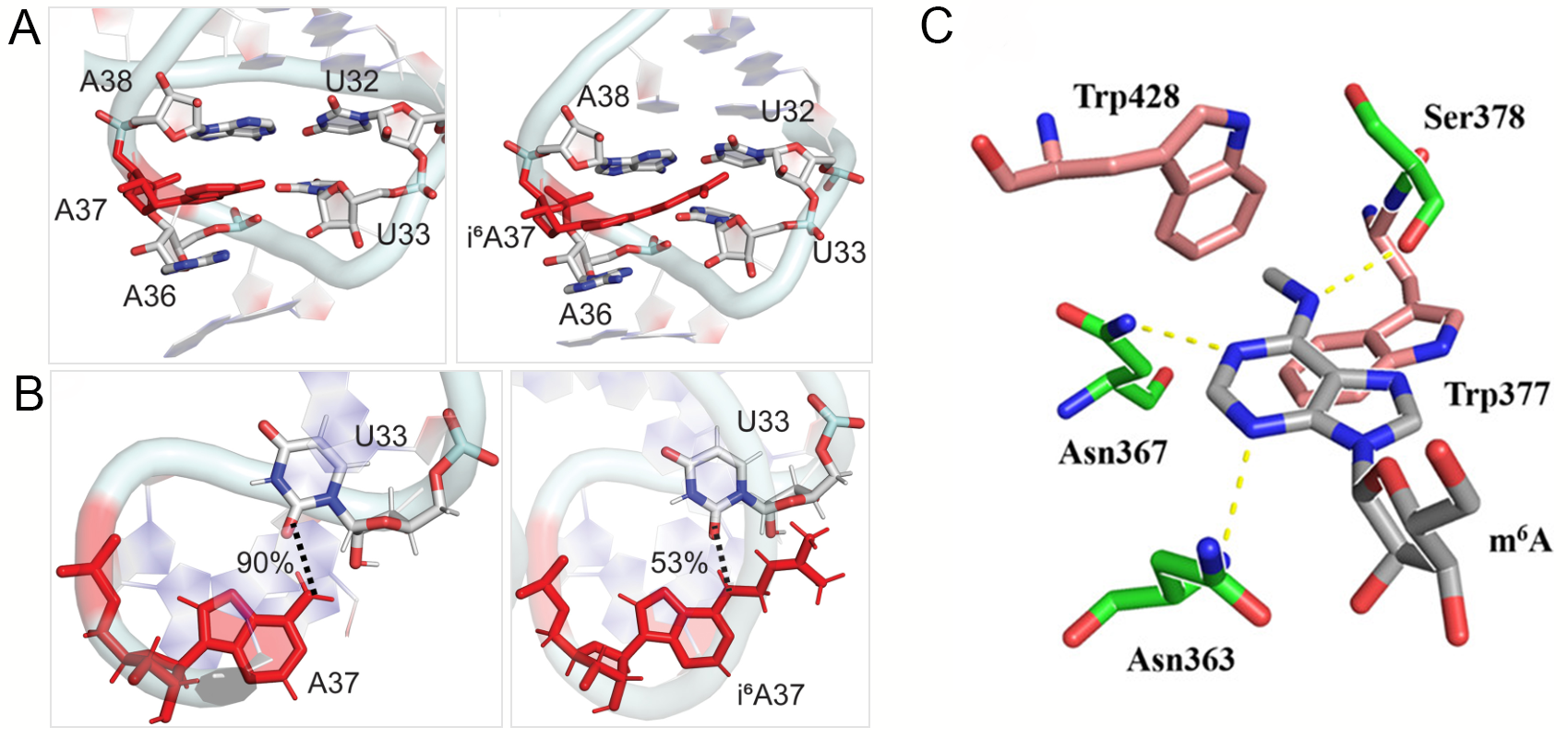}
\end{center}
\caption{
\small
(A) Stacking interactions between unmodified A37 and modified i$^6$A37 and neighboring nucleobases in the anticodon loop.
(B) Hydrogen-bonding interactions (with occupancies) between base A37/i$^6$A37 and Watson-Crick base U33 as obtained with MD simulations.
Adapted with permission from \cite{prabhakar2021posttranscriptional}
(C) Detailed interactions between m$^6$A and pocket of YTHDC1.  Adapted with permission from \cite{li2019flexible}.
The full structure of this complex can be seen in Fig.~\ref{fig-alchemistry}.
\label{fig-applications}
}
\end{figure}

A number of works
used MD simulations to investigate the structural role of  modifications on entire tRNAs,
suggesting them to be crucial for the stabilization of the functional structure.
A comprehensive study on 3 tRNAs was reported by \cite{zhang2014influence}, comparing all-modified with nonmodified tRNAs.
Modifications were shown to increase the rigidity of the anti-codon stem-loop, presumably facilitating pairing with mRNA during translation.
Overall, the effect of the modifications was suggested to be non-trivial, making tRNA more rigid in some regions and more flexible in other regions.
Similar results were obtained by \cite{xu2016structural}.
The effect of single modifications was analyzed by \cite{sonawane2016comparative} (G vs m$^2$G and m$^2$$_2$G)
and by \cite{prabhakar2021posttranscriptional} (G or A vs a total of 9 modifications),
both showing that a single modified position
could crucially stabilize the functional structure of the anticodon loop (see Fig.~\ref{fig-applications}A and B).

Other works directed the effort on the simulation of the anticodon stem loop.
\cite{galindo2016probing} showed that a hypermodified nucleotide stabilized interaction with a model mRNA fragment.
\cite{sambhare2014structural} studied the dynamics of the hypermodified nucleotide,
\cite{xiao2016simulation} discussed how this modification modulated the interaction with a designed peptide,
and \cite{vangaveti2022physical} analyzed the interaction with insulin mRNA, specifically
monitoring how the modification tuned the van-der-Waals interactions and the hydration of the stem loop.
The effect of anticodon modifications was also studied by \cite{sambhare2014structural} and by \cite{sonawane2015influence}.
\cite{vangaveti2020structural} reported how modifications at the wobble position of the anticodon loop
affect pairing with mRNA and codon recognition.
Finally, \cite{wang2016synthesis} presented a synthesis protocol for geranylated nucleotides.
Their work also includes MD simulations of geranylated nucleotides 
in RNA and DNA duplexes and in anti-codon--codon pairs, including the ribosomal subunit with all its associated proteins,
showing how geranylation of udirine affects U{\textperiodcentered}G pairing.

\subsubsection{Molecular simulations of entire ribosomes.}

Ribosomes contain a large number of post-transcriptional modifications. Therefore, MD simulations
of the ribosome usually  take these modifications into account. %
However, in many cases the modifications are not directly involved in the
process of interest.
\cite{pavlova2017toward} performed simulations of the \emph{E. coli} ribosome in multiple variants to characterize the relationship
between mutations and modifications and resistance to macrolides, a widely prescribed class of antibiotics.
Several mutations and modifications of an adenine in the exit tunnel were simulated (G, m$^6$A and m$^6$$_6$A) in complex with a number of macrolides.
The mutations and modifications were found to weaken the interaction of the ribosome with the macrolides, thus providing a microscopic explanation
for the observed antibiotic resistance.

\subsubsection{Molecular simulations of model systems and of other post-transcriptionally modified RNAs.}

A number of papers focused on the effect of post-transcriptional
modifications on the energetics and structure of other systems, ranging from individual base pairs to duplexes.

Some works addressed the dynamics of individual or paired nucleotides.
\cite{bavi2013md} reported short MD simulations of modified nucleotides (m$^2$G and m$^2$$_2$G) and compared them with crystal structures.
\cite{vendeix2009free} performed umbrella-sampling simulations to characterize the effect of modifications on base-pairing free energy.
\cite{hopfinger2020predictions} computed the thermodynamic stability of multiple modified nucleotides
using a combination of MD simulations and quantum mechanical calculations.

Several works specifically studied pseudouridine ($\Psi$).
A few of them addressed the impact of U to $\Psi$ modification in the structure and stability
of duplexes containing CUG or C$\Psi$G repeats.
\cite{delorimier2014modifications} reported enhanced-sampling calculations to estimate the free-energy
change associated to the opening of the central base pair in the repeat, which was found to be larger in the presence of one or two $\Psi$s,
relating the result to a change in the water coordination of the modified nucleotide.
A later work from the same group \citep{delorimier2017pseudouridine}  showed that the rigidity of $\Psi$ in the context of CUG and CCUG
repeats resulted in lower affinity with a RNA--binding protein.
Similar results were obtained by \cite{deb2019computational} using a different force field and thereby confirmed by NMR experiments.
Another modification of uridine, s$^2$U, has been studied by \cite{sarkar2020ensemble},
where it was shown that intercalation of the sulfur atom can induce order in single stranded RNAs with consecutive uridines.

Another commonly found modification is m$^6$A.
\cite{hurst2021deciphering}, after validating the employed force field using alchemical calculations on duplexes (see Section \ref{sec-validation}),
computed the free-energy landscape of a hairpin composed of a tetraloop and three base pairs with or without modifications, including m$^6$A.
Simulations were accelerated using temperature replica-exchange MD.
Remarkably, methylations at different positions were shown to stabilize or destabilize the hairpin structure in specific contexts,
which may provide a tool for RNA nanotherapeutic design.

The consequences of N1 adenosine and guanosine methylation (m$^1$A and m$^1$G) on duplex dynamics was studied by \cite{zhou2016m1a}.
In particular, this work shows that this modification changes the energetic balance between Watson--Crick
and Hoogsteen pairing, with the result of being not tolerated in RNA, where it can induce duplex melting.
We note that, compared with other common nucleobase methylations, m$^1$A shifts the nucleotide charge by +1, and that the destabilization of a RNA duplex by m$^1$A has been
reported to be sufficient to induce a completely different secondary structure in a tRNA \citep{voigts2007}.

A few papers addressed inosine (I).
Given the similarity in the hydrogen bond pattern formed by I and G, I can pair with both U and C, though the I{\textperiodcentered}C pair
only forms two hydrogen bonds.
\cite{krepl2013effect} studied the thermodynamics of I{\textperiodcentered}C pairs using MD simulations of RNA and DNA duplexes
subjected to alchemical calculations. The thermodynamic cycle was performed converting a G{\textperiodcentered}C pair
to a I{\textperiodcentered}C pair and computing the corresponding destabilization of the duplex, and results were correlated with experimental thermodynamic data.
\cite{sakuraba2020free} reported a more systematic study including more sequences as well as new experimental data,
confirming that MD can correctly predict differential stability of G{\textperiodcentered}C and I{\textperiodcentered}C pairs.
Less stable I{\textperiodcentered}U pairs were investigated using quantum mechanical calculations by \cite{jolley2015computational}.
\cite{vspavckova2018role} reported MD simulations of I{\textperiodcentered}U pairs,
showing that dynamics of tandem I{\textperiodcentered}U pairs depends on the neighboring base, with UII being
the most rigid sequence, with a limited impact on structure with respect to unmodified nucleotides.

Finally, not only tRNA and rRNA can be modified, as discussed in previous Sections, but also mRNA.
Simulations of tRNA in complex with portions of rRNA and mRNA
\citep{elliott2019modification} showed that 
2$^\prime$-O-methylation in mRNA
could hinder interactions that
are crucial during translation.

\subsection{Effects of post-transcriptional modifications on RNA recognition}

\subsubsection{RNA recognition by reader domain YTHDC1.}

The YTHDC1 is one of most studied m$^6$A readers. The RNA--protein complex has been solved for different oligonucleotide sequences, invariably
showing that m$^6$A is captured in an aromatic cage, with the flanking nucleotides laying at the RNA--protein surface.
Several MD studies have characterized the hydrogen-bond networks formed in the complex
\citep{li2019flexible,li2021atomistic,li2022structure,krepl2021recognition,zhou2022specific}.

\cite{li2019flexible} reported simulations with the CHARMM force field using a 5-nucleotides RNA single strand (5$^\prime$-GG(m$^6$A)CU-3$^\prime$),
highlighting the role of two tryptophan residues in the pocket, respectively stacking on m$^6$A and stabilizing its methyl group
(see Fig.~\ref{fig-applications}C).
The role of the flanking nucleotides was studied by both analyzing their fluctuations in the simulations and performing experiments with shorter variants.
\cite{li2021atomistic} continued this work by performing alchemical calculations to estimate the complex stabilization induced by
m$^6$A, using as a reference an isolated nucleoside. Stabilization was slightly overestimated with respect to reference experimental data.
A crucial water molecule was identified in the binding site and also studied with alchemical methods.

\cite{krepl2021recognition} published a related study with a similar RNA sequence (5$^\prime$- CG(m$^6$A)CAC-3$^\prime$) using AMBER parameters and specifically
developed charges. The authors performed alchemical calculations, identifying a water molecule entering the binding site
in a position occupied by the methyl group.
The stabilization of the complex was overestimated when compared to experiment, as in \citep{li2021atomistic}.
Results were shown to be dependent on the protocol used to initialize the alchemical calculation.

\subsubsection{Other RNA--protein interactions.}

\cite{gonzalez2020computational} characterized the interaction between RNA strands with a single oxidized G (8-oxoG) and a 
polynucleotide phosphorylase implicated in RNA turnover. Interestingly, they developed a protocol based on relatively short MD simulations
to assess the affinity of mutated protein sequences, validating the optimized sequences by affinity measurements.

\subsection{Molecular dynamics simulations of synthetic nucleic acids}

Synthetic nucleic acids often display backbone modifications that are introduced to facilitate their therapeutic use,
often aimed at modulating degradation and/or affinity with the target sequences.
MD simulations have been used to understand how the modifications impact backbone flexibility and hybridization energies.
\cite{gore2012synthesis} synthesized siRNAs including a 4$^\prime$-C-aminomethyl-2$^\prime$-O-methyl modification.
They then used MD simulations to show that the modification induces changes in the sugar puckering,
flexibility in groove dimension, and disturbances in hydrogen bonding and base stacking,
resulting in lower duplex stability as confirmed experimentally.
In a later work, \cite{harikrishna2017probing} used plain MD simulations of siRNA in complex with Argonaute 2
to elucidate how a number of synthetic modifications affect protein--RNA interactions.
Simulations were performed on the microsecond timescale, showing that the minimum timescale to see this conformational
variability is of the order of 300 ns.
\cite{masaki2010linear} characterized a number of 2$^\prime$-O-modified nucleotides.
These modifications affect the sugar flexibility and thus the fluctuations of duplex helical parameters.
The authors identified linear relationships between the predicted fluctuations and the duplex thermal
stability, suggesting  the possibility of predicting the thermal stability of 2$^\prime$-O-modified duplexes at the computer-aided molecular design stage.
This idea was then generalized to other modifications, also covering the nucleobase \citep{masaki2012prediction}.

\cite{seio2012short} developed a modified nucleotide that can be used at the 5$^\prime$ termini of oligonucleotides so as to
distinguish complementary polynucleotide depending on their length, for instance to distinguish microRNAs and pre-microRNAs.
In this work, they used MD simulations to construct structural models of the resulting double helix and to  
characterize the interaction of the modified nucleotide with the terminal phosphate of the recognized RNA.

Phosphorothioate (PT) modification of RNA backbone is generally considered an artificial modification,
though it has been recently reported to occur naturally. \cite{zhang2021phosphorothioate} studied the effect of
PT in a riboswitch in which the modification was artificially included and compared
the resulting trajectories with NMR data. The authors observed that existing force-field parameters cannot recapitulate the interactions seen in experiment,
and suggested that polarizable force fields might be necessary to describe this modification.
Interestingly, \cite{jing2019molecular} proposed an extension of the AMOEBA polarizable force field
to a number of modified nucleotides, including PT. Alchemical calculations were used to compute changes in duplex hybridization free energy
induced by the modification, reporting results in general agreement with experimental data.

Other interesting synthetic backbone modifications are locked nucleic acids (LNAs). Simulations with these modifications have been mostly conducted on LNAs within DNA or fully modified sequences
and are thus not covered here.

\section{Discussion and challenges ahead}

In this Chapter we reviewed selected applications of molecular dynamics (MD) simulations to modified RNAs,
covering validation of force-field parameters,
effects of modifications on dynamics or recognition, and synthetic nucleotides.
A recent related review complements ours, by providing a different perspective \citep{d2022challenges}.
A number of issues can be highlighted from our survey.

MD simulations of non-modified RNAs are routinely performed by many groups,
thus providing a significant mass of reference work.
On the contrary, simulations of modified nucleotides are sparse, with
most modifications simulated in a handful of papers. Hence,
force fields has not been validated to the same extent.
Researchers approaching the field should carefully validate parameters and 
be ready to develop new ones.
Integration of experimental data
\citep{bernetti2022combining} could provide a significant step forward.

Advanced sampling techniques are used rarely in this field.
However, flexible and structurally heterogeneous RNA molecules require enhanced
sampling methods to be faithfully characterized.
Possibly, this shortfall is a consequence of the fact that being able to carefully
design studies on
systems with complex modified nucleotides require a deep biochemistry knowledge that is not commonly available in the
community using enhanced sampling methods, and vice versa. Collaborative works could open the way to
the application of state-of-the-art simulation methods in this field.

Nevertheless, the presented applications show that MD simulations are at the level of
being useful in the interpretation and design of experiments. Given the growing biological relevance of
RNA modifications, we are confident that sinergy between simulation and experiment will become even stronger in the 
coming future.

\section{Acknowledgement}

Zhengyue Zhang and Stefano Bosio are acknowledged for reading a draft of this Chapter and providing useful suggestions.

\bibliographystyle{abbrvnat}
\bibliography{main}

\end{document}